\documentclass[pre,twocolumn,showpacs,superscriptaddress,floatfix]{revtex4}
\usepackage[latin1]{inputenc}
\usepackage{graphicx}
\usepackage{amsmath}
\usepackage{amssymb}
\usepackage{color}

\begin{document}

\title{The effective slip-length tensor for a flow over ``weakly'' slipping  stripes}

\author{Evgeny S. Asmolov}
\affiliation{A.N.~Frumkin Institute of Physical
Chemistry and Electrochemistry, Russian Academy of Sciences, 31
Leninsky Prospect, 119991 Moscow, Russia}
\affiliation{Central Aero-Hydrodynamic Institute, 140180
Zhukovsky, Moscow region,  Russia}
\affiliation{Institute of Mechanics, M. V. Lomonosov Moscow State University, 119071 Moscow, Russia}

\author{Jiajia Zhou}
%\email[]{zhou@uni-mainz.de}
\affiliation{Institut f\"ur Physik, Johannes Gutenberg-Universit\"at Mainz, 55099 Mainz, Germany}

\author{Friederike Schmid}
%\email[]{friederike.schmid@uni-mainz.de}
\affiliation{Institut f\"ur Physik, Johannes Gutenberg-Universit\"at Mainz, 55099 Mainz, Germany}

\author{Olga I. Vinogradova}
\affiliation{A.N.~Frumkin Institute of Physical
Chemistry and Electrochemistry, Russian Academy of Sciences, 31
Leninsky Prospect, 119991 Moscow, Russia}
\affiliation{Department of Physics, M.V.~Lomonosov Moscow State University, 119991 Moscow, Russia }
\affiliation{DWI, RWTH Aachen, Forckenbeckstr. 50, 52056 Aachen, Germany}
%\date{\today}

\begin{abstract}

We discuss the flow past a flat heterogeneous solid surface decorated by slipping stripes.
The  spatially varying slip length, $b(y)$, is assumed to be small compared to the scale of the heterogeneities, $L$, but finite. For such  ``weakly'' slipping surfaces, earlier analyses have predicted that the effective slip length is simply given by the surface-averaged slip length, which implies that the effective slip-length tensor becomes isotropic.
Here we show that a different scenario is expected if the local slip length has step-like jumps at the edges of slipping heterogeneities. In this case, the next-to-leading term in an expansion of the effective slip-length tensor in powers of $\mbox{max}\,(b(y)/L)$ becomes comparable to the leading-order term, but anisotropic, even at very small $b(y)/L$. This leads to an anisotropy of the effective slip, and to its significant reduction compared to the surface-averaged value.
The asymptotic formulae are tested by numerical solutions and are in agreement with results of dissipative particle dynamics simulations.
\end{abstract}

\maketitle

\section{Introduction}

With emerging technologies in microfluidics~\cite{stone2004,squires2005},
there has been renewed interest in quantifying the effects of surface
chemical heterogeneities with local scalar slip~\cite{vinogradova1999,bocquet2007} on fluid motion.
Well-known examples of such heterogeneous systems include composite superhydrophobic (Cassie) surfaces, where a gas layer is stabilized by a rough wall texture~\cite{quere.d:2008}. These surfaces are known to be
self-cleaning and show low adhesive forces. In addition, they also exhibit drag reduction for fluid
flow~\cite{bocquet2007,mchale.g:2010,vinogradova.oi:2011,vinogradova.oi:2012}. This is due to a local slip length at the gas areas, $b \simeq e (\mu/\mu_g - 1) \simeq 50 e$, where $\mu_g$ and $\mu$ are dynamic viscosities of a gas and a liquid, and $e$ is the thickness of the gas layer~\cite{vinogradova.oi:1995a}. As a result, and in contrast to a smooth hydrophobic surfaces, where $b$ cannot exceed a few tens of nm~\cite{vinogradova.oi:2003,vinogradova.oi:2009,charlaix.e:2005,joly.l:2006}, slip lengths up to tens or even hundreds of $\mu$m may be obtained for superhydrophobic textures~\cite{choi.ch:2006,joseph.p:2006}. Therefore, these surfaces have the potential to
influence microfluidics (or to extend microfluidic systems to nanofluidics),
by generating very fast and well-controlled flows in small devices~\cite{vinogradova.oi:2011,rothstein.jp:2010,vinogradova.oi:2012}.

\begin{figure}[tbp]
\begin{center}
\includegraphics[width=0.7\columnwidth]{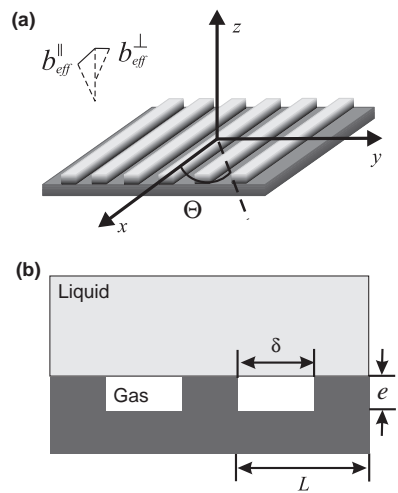}
\end{center}
\caption{Sketch of the striped surface: $\Theta=\protect\pi/2$ corresponds
to transverse stripes, $\Theta=0$ to longitudinal stripes (a), and of the liquid interface in the Cassie state (b). }
\label{fig:geometry}
\end{figure}

In case of superhydrophobic materials it is convenient to construct so-called effective slip
boundary conditions, where the complex flow pattern at a heterogeneous surface is replaced by an effective flow averaged over the length scale of the experimental configuration~\cite{vinogradova.oi:2011,kamrin_etal:2010}.
In other words, rather than trying to solve equations of motion on the scale of the individual corrugation or pattern, one considers the ``macroscale'' fluid motion (on the scale larger than the pattern characteristic length) by using macroscopically equivalent boundary conditions for an imaginary smooth surface.
Such an effective condition mimics the actual one along the true heterogeneous surface.
It fully characterizes the flow at the real surface and can be used to solve complex hydrodynamic problems with much reduced computational effort. The effective slip approach has been supported by statistical diffusion arguments~\cite{Bazant08}, and justified for the case of Stokes flow over a broad class of surfaces~\cite{kamrin_etal:2010}. Several numerical approaches have recently confirmed the concept of effective slip either at the molecular scale, using molecular dynamics~\cite{priezjev.n:2011,tretyakov.n:2013}, or at larger mesoscopic scales using finite element methods~\cite{priezjev.nv:2005,cottin:2004}, Lattice-Boltzmann~\cite{schmieschek.s:2012,asmolov.es:2013} or Dissipative Particle Dynamics~\cite{zhou.j:2012} simulations.

For an anisotropic texture, the effective boundary condition generally depends on the direction of the flow and is a tensor, $\mathbf{b}_{\mathrm{eff}}\equiv \{b_{\mathrm{eff}}^{ij}\}$, represented by a symmetric, positive definite $2\times 2$ matrix, {which can be} diagonalized by a rotation with angle $\Theta $ (Figure~\ref{fig:geometry}).
For all anisotropic surfaces its eigenvalues $b_{\mathrm{eff}}^{\parallel }$ and $b_{\mathrm{eff}}^{\perp}$ correspond to the fastest (greatest forward slip) and slowest (least forward slip) orthogonal directions~\cite{Bazant08}.
In the general case of arbitrary direction $\Theta$, the flow past such surfaces with anisotropic effective slip becomes misaligned with the driving force.
Therefore, anisotropic textures can potentially be used to generate transverse hydrodynamic flow~\cite{ajdari2002,Bazant08,stroock2002b}, which is of obvious fundamental and practical interest. For example, transverse hydrodynamic couplings in flow through a textured channel can be used to separate/concentrate suspended particles~\cite{gao.c:2008} or for passive chaotic mixing~\cite{stroock2002b,stroock2002a}.
This can also be used to generate anisotropic electrokinetic flows~\cite{bahga:2009,belyaev.av:2011a,Squires08}.

However, it has been predicted that regardless of the anisotropy of the surface texture, the effective slip-length tensor, $\mathbf{b}_{\mathrm{eff}}$, becomes isotropic ($b_{\mathrm{eff}}^{\perp }=b_{\mathrm{eff}}^{\parallel }$) for a \emph{``weakly''} slipping pattern, i.e. when the local
slip length, $b(x,y)$, is small compared to the characteristic scale of heterogeneities, $L$. The value of the effective slip is the surface average of the local slip length, $\mathbf{b}_{\mathrm{eff}}=\mathbf{I}\,\langle b(x,y)\rangle$.
In the particular case of a no-slip plane covered by patterns with constant slip length $b$ {-- the situation considered in most previous publications on the subject ~\cite%
{priezjev.nv:2005,ybert.c:2007,feuillebois.f:2009} --} one can {derive%
}~\cite{ybert.c:2007,Belyaev:2010,kamrin_etal:2010}
\begin{equation}
b_{\mathrm{eff}}^{\parallel ,\perp }\simeq b\phi ,  \label{average}
\end{equation}%
where $\phi =\delta /L$ is the surface fraction of the slipping phase. We remark and stress that $b_{\mathrm{eff}}^{\parallel ,\perp }$ can still remain extremely large compared to the nanometric scalar slip at flat
hydrophobic solids.
Eq.~(\ref{average}) implies, among other, that the flow aligns with the applied
driving force for all in-plane directions.
Thus, it seems impossible to
generate transverse hydrodynamic~\cite{vinogradova.oi:2011,zhou.j:2012} or transverse electro-osmotic~\cite{belyaev.av:2011a} phenomena for ``weakly''
slipping anisotropic textures.
Another important, and somewhat remarkable,
consequence of Eq.~(\ref{average}) is that the effective slip {is predicted
to depend} only on the fractions of slipping areas, but not on their
detailed structure.

Known derivations of Eq.~(\ref{average}), however,
neglect localized flow perturbations around possible jumps in discrete slip
lengths, from 0 to $b$, at the border of heterogeneities. Such jumps could contribute to the friction, as has been recently detected in a molecular dynamics simulation study~\cite{tretyakov.n:2013}, and also to the anisotropy of the flow, but we are not
aware of any prior work that has quantified the phenomena. In this paper we reconsider the problem of flow past ``weakly'' slipping one-dimensional surfaces, focusing on the situation of superhydrophobic stripes, where the perturbation of $b(y)$ is piecewise constant, i.e., it jumps in a step-like fashion at heterogeneity boundaries.

Our paper is arranged as follows: In Sec.~\ref%
{sec:theory} we define the problem
and construct the expansions for {the}
eigenvalues of the slip-length tensor of alternating ``weakly'' slipping
stripes. Here we also analyze a singularity of {the} velocity gradient at
the edges of stripes. The details of the computer simulation method (dissipative particle dynamics) related to ``weakly'' slipping surfaces are discussed in Sec.~\ref{sec:simulation}.
Finally, in Sec.~\ref{sec:results}, we present simulation and numerical
results to validate the predictions of the asymptotic theory. The practical
implications and limitations of our models are also reviewed here.
In Appendix~\ref{appA} we give some simple arguments showing that standard two-term expansions for effective slip lengths of one-dimensional textures could not be applied in case of a discontinuous local slip.

\section{Theory}
\label{sec:theory}

\subsection{Problem Set-up}
\label{sec:setup}

We consider a creeping flow along a {planar} anisotropic wall, and a
Cartesian coordinate system $(x,y,z)$ (Figure \ref{fig:geometry}). The
origin of coordinates is placed at the flat interface, a one-dimensional texture varies
over a period $L$. Our analysis is based on the limit of a thick channel or
a single interface, so that the velocity profile sufficiently far above the
surface may be considered as a linear shear flow. Dimensionless variables
are defined {in terms of the reference length scale} $L$, the {asymptotic}
shear rate far above the surface, $G,$ and the fluid kinematic viscosity, $%
\nu .$

For a one-dimensional texture there exists a simple relation between
longitudinal and transverse effective slip lengths~\cite{asmolov:2012}
\begin{equation}
b_{\mathrm{eff}}^{\perp }[b\left( y\right) ]=\frac{b_{\mathrm{eff}%
}^{\parallel }\left[ 2b\left( y\right) \right] }{2},  \label{1D}
\end{equation}%
which has recently been verified for cosine variation in local slip length by using lattice Boltzmann simulations~\cite{asmolov.es:2013}.
Therefore, it is sufficient to consider the longitudinal configuration.
Since {in this case}, the velocity has only one component, we seek {a}
solution for the velocity profile {of} the form
\begin{equation*}
v=U+u,
\end{equation*}%
where $U=z\ $is the undisturbed linear shear flow. The perturbation of the
flow $u\left( y,z\right) $, which is caused by the presence of the texture
and decays far from the surface at small Reynolds number $Re=GL^{2}/\nu $,
satisfies the dimensionless Laplace equation,%
\begin{equation}
\Delta u=0.  \label{1.6}
\end{equation}%
The boundary conditions at the wall and at infinity are defined as%
\begin{eqnarray}
z=0: &\quad &u-\varepsilon \beta \left( y\right) \partial _{z}u=\varepsilon
\beta \left( y\right) ,\   \label{bcu} \\
z\rightarrow \infty : &\quad &\partial _{z}u=0\mathbf{,}  \label{bci}
\end{eqnarray}%
where $\varepsilon =\mbox{max}\,(b(y))/L\ $and $\beta =b\left( y\right) /\mbox{max}\,(b(y))$ is the
normalized slip length.

The solution {of Eqs.\ (\ref{1.6})-(\ref{bci}) }for a ``weakly'' slipping
anisotropic texture, $\varepsilon \ll 1$, can be constructed as {an
expansion in powers of $\varepsilon $ }:%
\begin{equation}
u=\sum\limits_{k=1}^{\infty }\varepsilon ^{k}\varphi _{k}.  \label{pow}
\end{equation}%
The boundary conditions to $\varphi _{k}$ can be readily obtained by
substituting Eq.\ (\ref{pow}) into Eq.\ (\ref{bcu}) and by collecting the
terms of the order of $\varepsilon ^{k}$~\cite{kamrin_etal:2010}:%
\begin{equation}
\begin{array}{rcl}
z=0: & \quad & \varphi _{1}=\beta \left( y\right) , \\
z=0,\ k>1: & \quad & \varphi _{k}=\beta \left( y\right) \partial _{z}\varphi
_{k-1}.%
\end{array}
\label{bc_kbs}
\end{equation}%
The leading-order solution yields an area-averaged isotropic slip length~%
\cite{ybert.c:2007,Belyaev:2010}. {In practice, this} means that the
slip-length tensor becomes isotropic and {that} for all in-plane directions,
the flow aligns with the applied force.

It is important to note however that Eqs.(\ref{pow}) and (\ref{bc_kbs}) are inapplicable for a discontinuous $\beta (y)$ (see Appendix~\ref{appA} for details).
{From a physical point of view, the
problem is associated with singularities of the velocity gradient at the
boundaries of the slip region.} As a specific example, let us {consider} a
classical case of alternating ``weakly'' slipping ($\varepsilon =b/L\ll 1$)
stripes {with}
\begin{equation}
\beta \left( y\right) =\left\{
\begin{array}{cl}
1 & \text{ as }\left\vert y\right\vert \leq \phi /2 \\
0 & \text{ as }\phi /2<\left\vert y\right\vert \leq 1/2%
\end{array}%
\right. ,  \label{by}
\end{equation}%
so that the boundary conditions, Eq.\ (\ref{bcu}), can be rewritten as%
\begin{equation}
\begin{array}{rcl}
z=0,\ \left\vert y\right\vert \leq \phi /2: & \quad & u-\varepsilon \partial
_{z}u=\varepsilon , \\
z=0,\ \phi /2<\left\vert y\right\vert <1/2: & \quad & u=0.%
\end{array}
\label{bc_str}
\end{equation}%
The velocity gradient grows infinitely near the edge of the slip region (see
Sec. \ref{sec:e_s} for a detailed analysis). As a result, the corresponding
term in Eq.\ (\ref{bc_str}), $\varepsilon \partial _{z}u$, {has the same
order of magnitude} as $u$ in {the} vicinity of the slipping boundary.
Therefore, it cannot be neglected {compared} to the leading order, even
though $\varepsilon $ is small.

\subsection{Slip-length tensor}
\label{sec:tensor}

We now consider the case of stripes more specifically. We first compute the
eigenvalues of the effective slip-length tensor. Since we assume only weak
local slippage, we evaluate {the} effective slip length in the principal
directions to second order in $\varepsilon$ and seek for a solution which is
finite, i.e., has no singularity.

A general solution satisfying the Laplace equation (\ref{1.6}) and decaying
at infinity can be presented in terms of {a} cosine Fourier series as~\cite%
{asmolov:2012}%
\begin{equation}
u=\frac{a^{0}}{2}+\sum\limits_{n=1}^{\infty }a^{n}\exp \left( -2\pi
nz\right) \cos \left( 2\pi ny\right) ,  \label{fu}
\end{equation}%
where $a^{n}$ are {constant coefficients} to be found from (\ref{bc_str}).
The Navier slip boundary condition (\ref{bcu}) can be written in terms of
the Fourier coefficients $a^{n},$ {accounting for} (\ref{fu}), as%
\begin{equation}
\frac{a^{0}}{2}+\sum_{n=1}^{\infty }\left[ 1+2\pi n\varepsilon \beta \left(
y\right) \right] a^{n}\cos \left( 2\pi ny\right) =\varepsilon \beta \left(
y\right) .\   \label{bc_ft}
\end{equation}%

{We} construct {the} asymptotic series for alternating stripes,%
\begin{equation*}
u=\sum\limits_{k=1}^{\infty }u_{k}=\sum\limits_{k=1}^{\infty }\left[ \frac{%
a_{k}^{0}}{2}+\sum\limits_{n=1}^{\infty }a_{k}^{n}\exp \left( -2\pi
nz\right) \cos \left( 2\pi ny\right) \right] ,
\end{equation*}%
imposing that $\left\vert u_{k+1}/u_{k}\right\vert \ll 1$ over the entire
flow region. The boundary conditions {for} $u_{k}$ at the wall can be chosen
as follows:%
\begin{equation}
z=0:\quad u_{k}-\varepsilon \partial _{z}u_{k}=r_{k}\left( y\right) ,
\label{bc_1}
\end{equation}%
\begin{equation}
r_{1}\left( y\right) =\left\{
\begin{array}{rl}
\varepsilon & \text{ as }\left\vert y\right\vert \leq \phi /2 \\
0 & \text{ as }\phi /2<\left\vert y\right\vert \leq 1/2%
\end{array}%
\right. ,  \label{bc_0}
\end{equation}%
\begin{equation}
k>1:\quad r_{k}\left( y\right) =\left\{
\begin{array}{cl}
0 & \text{ as }\left\vert y\right\vert \leq \phi /2 \\
-\varepsilon \partial _{z}u_{k-1} & \text{ as }\phi /2<\left\vert
y\right\vert \leq 1/2%
\end{array}%
\right. .  \label{bc_k}
\end{equation}%
The reader may check by the summation of Eqs.\ (\ref{bc_1}) over $k$ that
they are fully equivalent to Eq.\ (\ref{bc_str}).

The slip velocity is the average velocity over the period:%
\begin{equation*}
u_{slip}=\sum\limits_{k=1}^{\infty }a_{k}^{0}/2.
\end{equation*}%
The boundary condition (\ref{bc_1}) can be rewritten in view of (\ref{bc_ft}%
) as%
\begin{equation*}
\frac{a_{k}^{0}}{2}+\sum\limits_{n=1}^{\infty }a_{k}^{n}\left( 1+2\pi
\varepsilon n\right) \cos \left( 2\pi ny\right) =r_{k}\left( y\right) .
\end{equation*}%
The coefficients $a_{k}^{n}$ are now determined using the inverse Fourier
transform:%
\begin{eqnarray}
a_{k}^{0} &=&2\int_{-1/2}^{1/2}r_{k}\left( y\right) dy,  \label{a_k} \\
n>0:\quad a_{k}^{n} &=&\frac{2}{1+2\pi \varepsilon n}\int_{-1/2}^{1/2}r_{k}%
\left( y\right) \cos \left( 2\pi ny\right) dy.  \notag
\end{eqnarray}%
{From} Eqs.\ (\ref{bc_0}) and (\ref{a_k}), we have to {leading order in $%
\varepsilon $}
\begin{eqnarray*}
a_{1}^{0} &=&2\varepsilon \phi , \\
n>0:\quad a_{1}^{n} &=&\frac{2\varepsilon \sin \left( \pi n\phi \right) }{%
\pi n\left( 1+2\pi \varepsilon n\right) }.
\end{eqnarray*}%
To find the second-order {terms we must} evaluate $r_{2}=\partial _{z}u_{1}$%
, which gives%
\begin{eqnarray}
\partial _{z}u_{1} &=&-\sum\limits_{n=1}^{\infty }a_{1}^{n}2\pi n\cos \left(
2\pi ny\right)   \notag \\
&=&-4\varepsilon \sum\limits_{n=1}^{\infty }\frac{\sin \left( \pi n\phi
\right) \cos \left( 2\pi ny\right) }{1+2\pi \varepsilon n}.
\end{eqnarray}%
The second-order slip velocity is then%
\begin{equation}
a_{2}^{0}=-4\int_{\phi /2}^{1/2}\varepsilon \partial _{z}u_{1}dy=-\frac{%
4\varepsilon ^{2}}{\pi }\sum\limits_{n=1}^{\infty }\frac{1-\cos \left( 2\pi
n\phi \right) }{n\left( 1+2\pi \varepsilon n\right) }  \label{a_02}
\end{equation}%
\begin{equation}
=\frac{4\varepsilon ^{2}}{\pi }\left\{ \ln \left( 2\pi \varepsilon \right)
-\gamma -\frac{1}{2}\ln \left[ 4\sin ^{2}\left( \pi \phi \right) \right]
\right\} +O\left( \varepsilon ^{3}\right) ,  \label{a_023}
\end{equation}%
where $\gamma =0.5772157...$ is Euler's constant. The series in Eq.~(\ref{a_02})
are very similar to those expected for a discontinuous $b(y)$ (see Eq.~(\ref{sum1}) of Appendix~\ref{appA}). They differ only by the factor {$%
(1+2\pi \varepsilon n)$} in the denominator of the first sum. This factor is
small at $n=1,$ but it {grows linearly with $n$ at large $n$, thus
ensuring}
convergence of the series. Note that the first logarithmic term in (\ref%
{a_023}) does not depend on the fraction of the slip regions, $\phi .$ This
term is associated with the flow singularities near {the} boundaries between
no-slip and slip regions (see Subsection~\ref{sec:e_s}), which are
responsible for additional viscous dissipation that {reduces} $b_{\mathrm{eff%
}}.$

Finally, for the longitudinal effective slip we obtain the following
expansion%
\begin{equation}  \label{b_par}
b_{\mathrm{eff}}^{\parallel }/L=\varepsilon \phi +\frac{2\varepsilon ^{2}}{%
\pi }\left\{ \ln \left[ \frac{\pi \varepsilon }{\sin \left( \pi \phi \right)
}\right] -\gamma \right\} +O\left( \varepsilon ^{3}\ln \varepsilon \right) ,
\end{equation}%
{from which we can derive the transverse effective slip using (\ref{1D})},
\begin{equation}  \label{b_perp}
b_{\mathrm{eff}}^{\perp }/L=\varepsilon \phi +\frac{4\varepsilon ^{2}}{\pi }%
\left\{ \ln \left[ \frac{2\pi \varepsilon }{\sin \left( \pi \phi \right) }%
\right] -\gamma \right\} +O\left( \varepsilon ^{3}\ln \varepsilon \right) .
\end{equation}

{To summarize}, we have here directly demonstrated that {Eq.\ (\ref{average}%
) must be applied} with care. On the one hand, Eqs.\ (\ref{b_par}) and (\ref%
{b_perp}) unambiguously show that {\ Eq.\ (\ref{average}) does indeed give
the correct first-order term of an expansion} for {the} eigenvalues of the
slip-length tensor, even in a case of alternating stripes. On the other
hand, {the higher order contributions may be nonanalytical in $\varepsilon$,
which may create complications}. In case of a {local slip which exhibits
step-like jumps} at the edge of heterogeneities, the {second-order terms} of
the expansions become of the order of $\varepsilon^2 \ln \varepsilon$ (in
contrast to $\varepsilon^2,$ which would be expected for continuously
varying local slip). Therefore, they can be comparable to the first-order
terms and cannot be ignored even at relatively small $\varepsilon$ (see
Section \ref{sec:results}). These terms are not only responsible for
anisotropy of the flow, but also (being negative) for an additional
dissipation.

\subsection{Edge singularity}

\label{sec:e_s}

We now describe {the} flow singularities near slipping heterogeneities in
more detail. For the flow over a surface with rectangular grooves, the shear stress is found to be singular near sharp
corners{, i.e.,} proportional to $r^{-1/3}$ for longitudinal and {to} $%
r^{-0.455}$, for transverse configurations~\cite{wang2003}. Here $r$ is the distance from
the corner. Following this approach, we now consider the flow in {the}
vicinity of the edge of our ``weakly'' slipping regions, by using polar
coordinates $(r,\theta )$ with the origin in $\left( y,z\right) =\left( \phi
/2,0\right) $. The no-slip and slip regions then correspond to $\theta =0\ $%
and $\theta =\pi$. The solution of the Laplace equation (\ref{1.6}) that
satisfies the no-slip boundary condition at $\theta =0$ is%
\begin{equation}
u=cr^{\lambda }\sin \left( \lambda \theta \right) .  \label{ur}
\end{equation}%
The velocity at the edge is finite provided $\lambda >0.$ The components of
velocity gradient are%
\begin{eqnarray}
\partial _{z}u&=&c\lambda r^{\lambda -1}\cos \left[ \theta \left( 1-\lambda
\right) \right] ,  \notag \\
\partial _{y}u&=&-c\lambda r^{\lambda -1}\sin \left[ \theta \left( 1-\lambda
\right) \right] .  \label{dzu}
\end{eqnarray}%
The velocity decays faster than its gradient as $r\rightarrow 0:$ $%
r^{\lambda }$ vs. $r^{\lambda -1}.$ Hence, in a small region $r\sim
\varepsilon ,$ the dimensionless shear rate $\varepsilon \partial _{z}u$ is
of the same order as $u,$ and cannot be ignored in the boundary condition (%
\ref{bc_str}) even though $\varepsilon \ll 1.$ Moreover, at smaller
distances, $r\ll \varepsilon ,$ the term $\varepsilon \partial _{z}u$
dominates over $u$, and the condition in this region becomes shear-free:
\begin{equation}
r\ll \varepsilon ,\ \theta =\pi :\quad \partial _{z}u=0.  \label{sh_f}
\end{equation}%
{The} last condition enables us to find $\lambda $. To satisfy Eq.\ (\ref%
{sh_f}) one should require, in view of (\ref{dzu}), $\lambda =1/2.$
Therefore, the velocity over the slip region is
\begin{equation}
r\ll \varepsilon ,\ \theta =\pi :\quad u=cr^{1/2},\quad \partial
_{y}u=-cr^{-1/2}/2,  \label{si_s}
\end{equation}%
where $c$ is a constant. The velocity gradient over the no-slip region
follows from (\ref{dzu}):%
\begin{equation}
r\ll \varepsilon ,\ \theta =0:\quad \partial _{z}u=cr^{-1/2}/2.
\label{si_ns}
\end{equation}
In other words, {the shear stress has a singularity at the edge}.

We remark that Eqs.\ (\ref{si_s}) and (\ref{si_ns}) are valid in a small
region, $r\ll \varepsilon,$ near a jump in the discrete local slip length,
from $0$ to a finite $b.$ {Therefore}, our asymptotic theory is {only valid}
provided that the fractions of the slip and no-slip regions are not too
small, $\phi \gg \varepsilon ,\ 1-\phi \gg \varepsilon .$ Otherwise, the two
edges of heterogeneities are close to each other, so that the {singular
regions} overlap.
%FS: I know what you mean, but the notion of ''overlapping singularities'' is kind of odd.
Note that {a} similar, $r^{-1/2}$, dependence of the velocity has been
obtained earlier for a no-slip surface decorated {with} perfect-slip stripes
\cite{philip.jr:1972,sbragaglia.m:2007,asmolov:2012}. A striking conclusion from our analysis is that such a singularity appears even at a very small slip at the gas area.

For the transverse flow one can use the relation between the velocity fields
for the two orientations~\cite{asmolov:2012}:
\begin{gather}
v=\frac{1}{2}\left( u_{d}+z\frac{\partial u_{d}}{\partial z}\right) ,\quad
w=-\frac{z}{2}\frac{\partial u_{d}}{\partial y},  \label{vw} \\
p=-\frac{\partial u_{d}}{\partial y},  \label{p}
\end{gather}%
where $u_{d}\left( y,z\right) =u\left[ y,z,2\varepsilon \beta \left(
y\right) \right] $ is the velocity field for the longitudinal pattern with
double local slip length {(cf. Eq.\ (\ref{1D}))}. %
%FS: Is it correct to refer to this Equation here?
{Hence} we conclude that at the wall%
\begin{equation}
z=0:\quad v=\frac{1}{2}u_{d},\quad \frac{\partial v}{\partial z}=\frac{%
\partial u_{d}}{\partial z}.  \label{v0}
\end{equation}
{From Eqs.\ (\ref{si_s}) - (\ref{p}), it also follows} that $\frac{\partial v%
}{\partial z},$ $\frac{\partial w}{\partial z}$ and $p$ all have the same
singularity $r^{-1/2}$ at the edge of ``weakly'' slipping region.

\section{Simulation method}
\label{sec:simulation}

We apply Dissipative Particle Dynamics (DPD) method \cite{Hoogerbrugge1992, Espanol1995, Groot1997} to simulate the flow near striped superhydrophobic surfaces.
The DPD method is an established coarse-grained, momentum-conserving method for mesoscale fluid simulations, which naturally includes thermal fluctuations.
More specifically, we use a DPD version without conservative interactions \cite{Soddemann2003}.
The hydrodynamic boundary conditions are implemented using the tunable-slip method \cite{Smiatek2008}, which model the fluid/surface interaction using an effective friction force, combined with an appropriate thermostat.

The error of the simulation data is obtained from averaging over six independent runs. The absolute error in the effective slip length is typically around $0.2\,\sigma$, where $\sigma$ is the length unit in the simulation (see Appendix~\ref{app:dpd} for details).
For ``weakly'' slipping surfaces, the ratio between the effective slip length and the stripe spacing is of the order of $b_{\rm eff}/L\sim 0.1$.
One can then improve the accuracy by choosing a large $L$.
On the other hand, the size of the simulation box is proportional to $L^2$, and the time for the system to reach a steady state also increases for a large system.
Therefore, the choice of the stripe spacing is a compromise between the computational accuracy and time.
In this study, we have used a stripe spacing of $L=100\sigma$ and a simulation box of size $20\sigma \times 100 \sigma \times 102 \sigma$.
With a density of $3.0\,\sigma^{-3}$, a typical system consists of $6\times 10^5$ particles.
The simulations are carried out using the open source simulation package
ESPResSo \cite{ESPResSo}.

Based on the values of the velocities close to the surface, we can estimate the characteristic Reynolds number of our system to be of $O(10)$, which is larger than in real microfluidic devices.
Thus, inertia effects may become important in simulations, and the Stokes equation is not strictly valid.
This leads to a slight reduction of our simulation results for the effective slip length transverse stripes as we discuss below.
To reach more realistic Reynolds numbers, we would need to reduce the shear rate by orders of magnitude.
This would reduce the average flow velocity significantly, and the necessary simulation time to gather data with sufficiently good statistics will then increase prohibitively.

%%%%%%%%%%%%%%%%%%%%%%%%%%%%%%%%%%%%%%%%%%%%%%%%%%%%%%%%%%%%%%

\section{Results and Discussion}
\label{sec:results}

In this section, we compare predictions of our asymptotic %continuous
theory with results of DPD simulations and {direct numerical solutions of
Eqs.\ (\ref{1.6})-(\ref{bci}).} To find $a^{n}$ numerically we truncate the
sum in (\ref{bc_ft}) at some cut-off number $N$ (usually $N=501$) and
evaluate it in the points $y_{l}=l/2\left( N-1\right) ,$ where $l$ are
numbers varying from $0$ to $N-1$. Then Eq.~(\ref{bc_ft}) is reduced to a linear
system $A_{n}^{l}a^{n}=\varepsilon \beta ^{l},$ where $A_{n}^{l}=\left[
1+2\pi n\varepsilon \beta \left( y_{l}\right) \right] \cos \left( 2\pi
ny_{l}\right) $ and $\beta ^{l}=\beta \left( y_{l}\right) .$ The system is
solved using the IMSL routine LSARG.

\begin{figure}[htbp]
\includegraphics[width=1.0\columnwidth]{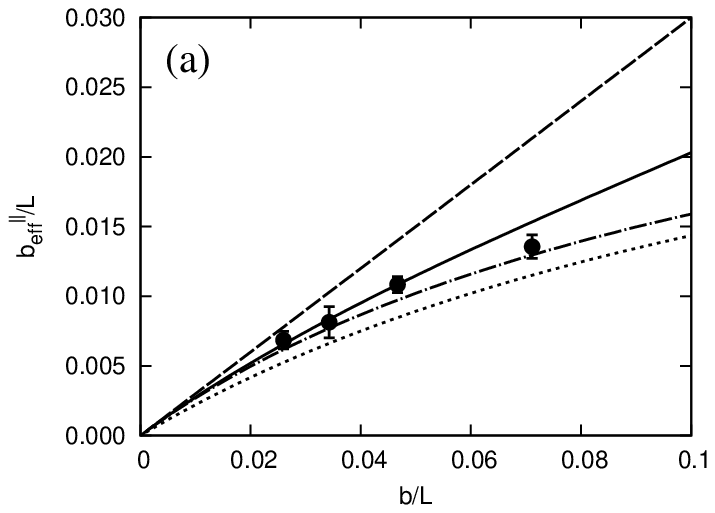} %
\includegraphics[width=1.0\columnwidth]{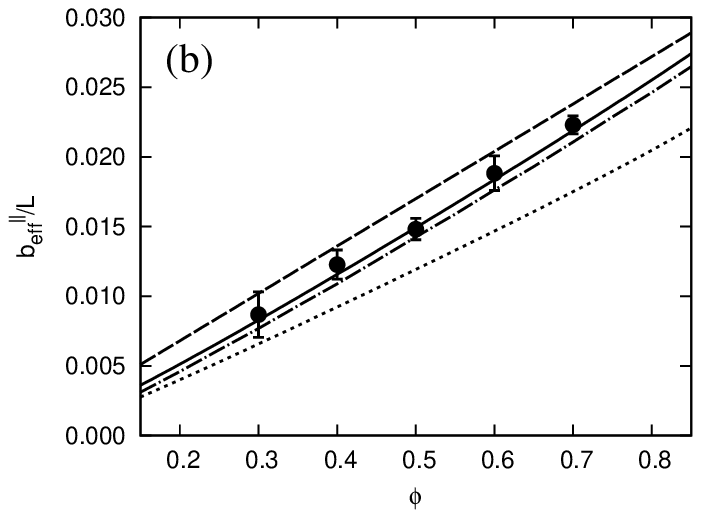}
%(b)  \includegraphics[width=0.5\columnwidth]{log3_ort.eps}
\caption{The longitudinal effective slip length as functions of {(a) the}
local slip for the texture with $\protect\phi=0.3$, and (b) {the} fraction
of the slipping phase at $b/L=0.034$. Symbols are simulation data.
Dash-dotted curves show exact numerical results, solid lines correspond to
the two-term logarithmic expansions (Eq.\ (\protect\ref{b_par})), dash lines
to the linear theory (Eq.\ (\protect\ref{average})), dotted lines to Eq.\ (%
\protect\ref{beff_par_largeH}).}
\label{fig:log_b}
\end{figure}

Figures~\ref{fig:log_b}(a) and (b) show the exact numerical results and DPD
simulation data for the longitudinal component of the slip-length tensor as {%
a function} of the dimensionless slip length $b/L$ and the slipping area $%
\phi$. The simulation data are in excellent agreement with {the} numerical
results, confirming the validity of our DPD scheme. Similar calculations {%
were} made for {the} transverse component of the slip-length tensor. All
curves were found to be very similar to {those} presented in Figures~\ref%
{fig:log_b}, {therefore, we do not show them here}. {The values for the
transverse component are smaller than those for the longitudinal component,
indicating that the flow is anisotropic. The simulation data in the
transverse case tend to be slightly smaller than the prediction from the
numerical solution. This has been observed previously \cite{zhou.j:2012} and
can be related to the relatively large Reynolds numbers in our system (see
Sec. \ref{sec:simulation}). Inertia effects influence the flow past
transverse stripes, as will be discussed below in the context of Figure~\ref%
{fig:slip_vel}. For flow past longitudinal stripes, $\mathbf{u}=\left( u\left( y,z\right)  ,0,0\right) $, the inertia effects are negligible, since convective terms
in the Navier-Stokes equations, $\left( \mathbf{u\cdot
\nabla }\right) \mathbf{u}=\mathbf{0}$. Thus the DPD data shown in Fig.\ \ref{fig:log_b} are not
affected by Reynolds number.}

The surface-averaged slip, predicted by Eq.\ (\ref{average}), is also shown
in Figure~\ref{fig:log_b} and is well above the exact values of the
longitudinal effective slip. Also included in Figure~\ref{fig:log_b} are the
predictions of our theoretical result, Eq.\ (\ref{b_par}). {One can see that
Eq.\ (\ref{b_par}) indeed gives the correct asymptotic behavior in the limit
of very small $b/L$. It slightly overestimates the value of the longitudinal
effective slip at larger $b/L$.} We remark and stress that nevertheless, our
second-order calculation is much more accurate than Eq.~(\ref{average}).

Recently, \cite{Belyaev:2010} suggested approximate expressions for
effective slip lengths of a surface decorated by partial slip stripes:
\begin{equation}  \label{beff_par_largeH}
b_{\mathrm{eff}}^{\parallel} \simeq \frac{L}{\pi} \frac{\ln\left[\sec\left(%
\displaystyle\frac{\pi \phi}{2 }\right)\right]}{1+\displaystyle\frac{L}{\pi b%
}\ln\left[\sec\displaystyle\left(\frac{\pi \phi}{2 }\right)+\tan\displaystyle%
\left(\frac{\pi \phi}{2}\right)\right]},
\end{equation}
\begin{equation}  \label{beff_ort_largeH}
b_{\mathrm{eff}}^{\perp} \simeq \frac{L}{2 \pi} \frac{\ln\left[\sec\left(%
\displaystyle\frac{\pi \phi}{2 }\right)\right]}{1+\displaystyle\frac{L}{2
\pi b}\ln\left[\sec\displaystyle\left(\frac{\pi \phi}{2 }\right)+\tan%
\displaystyle\left(\frac{\pi \phi}{2}\right)\right]}.
\end{equation}
These formulae have been verified~\cite{Belyaev:2010} using the method
developed by~\cite{cottin:2004}. The agreement between the theoretical and
numerical data was found to be very good for all $\phi$ and $b/L$, but at $%
b/L=O(1)$, {small discrepancies were observed}, suggesting that Eqs.\ (\ref%
{beff_par_largeH}) and (\ref{beff_ort_largeH}) slightly underestimate the
effective slip {length}. To examine this more closely, we {also} include {%
the prediction of} Eq.\ (\ref{beff_par_largeH}) in Figure~\ref{fig:log_b}.
We {find} indeed a small discrepancy between {the} exact numerical data and
the predictions of Eq.\ (\ref{beff_par_largeH}), which gives smaller {values
for the slip length}. The same trends were observed in a wide range of $\phi$%
, and the discrepancy slightly increases with the fraction of slipping
phase. Still, {the} analytical expressions for the effective slip by~\cite%
{Belyaev:2010} appear to be surprisingly accurate, {given} their simplicity.
We stress, however, that they do not {reproduce the} asymptotic result, Eq.\
(\ref{average}), in the limit of very small $b/L$. They do correctly predict
a linear dependence on $b$ in the limit of ``weakly'' slipping stripes
\begin{equation}
b_{\mathrm{eff}}^{\parallel }=b_{\mathrm{eff}}^{\perp }\simeq f(\phi) \, b,
\label{beff_smallb}
\end{equation}
but the prefactor, $f(\phi)$, differs from $\phi $:
\begin{equation}
f(\phi)\simeq\frac{\ln \left[ \sec \left( \frac{\pi \phi }{2}\right) \right]
}{\ln \left[ \sec \left( \frac{\pi \phi }{2}\right) +\tan \left( \frac{\pi
\phi }{2}\right) \right] }< \phi  \label{factor}
\end{equation}%
This prefactor corresponds to the slope of the curve $b_{\mathrm{eff}%
}^{\parallel }\left( b\right) $ at $b/L=0$. {In Figure \ref{fig:log_b}}, the
slope of the dotted line {corresponding to Eq.\ (\ref{beff_par_largeH})} is
smaller than the exact one. {Nevertheless}, the values of $b_{\mathrm{eff}%
}^{\parallel }$ and $b_{\mathrm{eff}}^{\perp }$ given by (\ref%
{beff_par_largeH}) and (\ref{beff_ort_largeH}) correlate well with the
numerical data for all $\phi $ and small but finite $b/L.$

\begin{figure}[tbp]
\includegraphics[width=1.0\columnwidth]{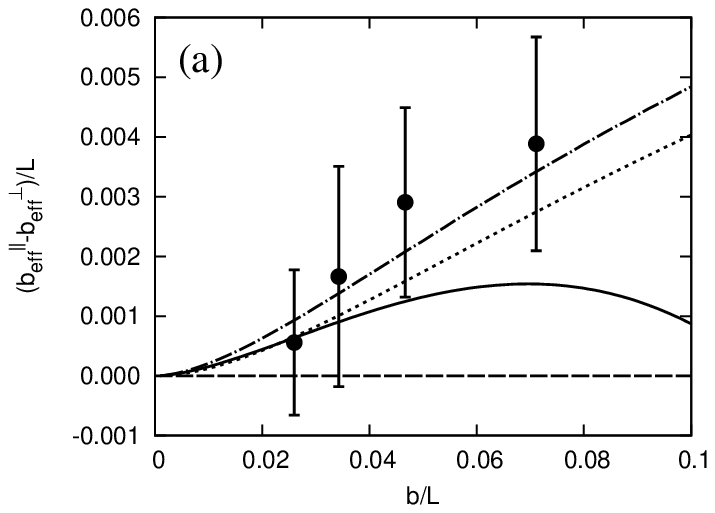} %
\includegraphics[width=1.0\columnwidth]{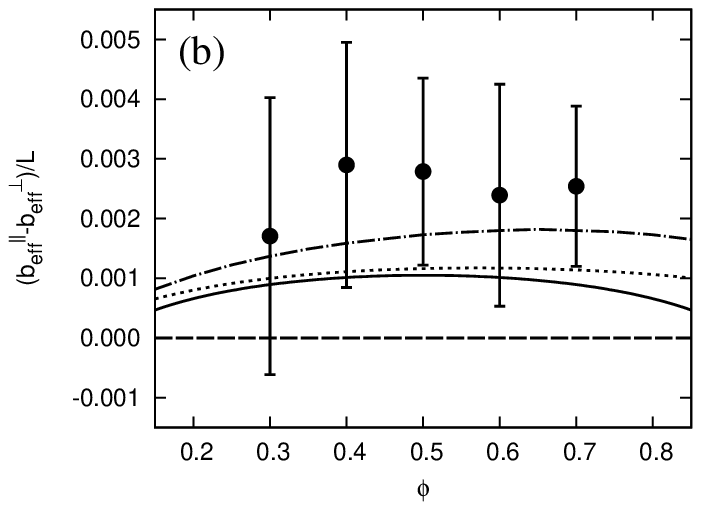}
\caption{The difference between longitudinal and transverse effective slip
lengths as functions of (a) the local slip for textures with $\protect\phi%
=0.3$, and (b) $\protect\phi$ for textures with $b/L=0.034$. Solid curves
correspond to calculations made with the two-term logarithmic expansions
(Eq.\ (\ref{b_dif})). Dotted curves are
obtained using Eqs.\ (\protect\ref{beff_par_largeH}) and (\protect\ref%
{beff_ort_largeH}). Other notations are the same as in Figure \protect\ref%
{fig:log_b}.}
\label{fig:log_phi}
\end{figure}

To summarize, ``weakly'' slipping stripes generate anisotropic effective slippage compared to simple, smooth
channels, an ideal situation for various potential applications. To illustrate this, we now show that our results may be
used to easily quantify transverse phenomena (important for a passive microfluidic mixing) and a
reduction of the hydrodynamic drag force.

We begin by discussing a transverse
flow, or the flow anisotropy, which in a thick channel has been predicted to be controlled by the difference between the
eigenvalues of the effective slip tensor, $b_{\mathrm{eff}}^{\parallel }-b_{%
\mathrm{eff}}^{\perp }$, which in turn depends on $\phi$ and $b$~\cite%
{vinogradova.oi:2011}. According to Eq.~(\ref{average}), this difference
should vanish for ``weakly'' slipping surfaces. The effect of anisotropy is
highlighted in Figure~\ref{fig:log_phi}(a) and (b), which shows the
difference between the longitudinal and transverse effective slip lengths
computed for fixed $\phi=0.3$ and $b/L=0.034$, respectively. The exact
numerical values are positive, except in {the} case of extremely small local
slip, clearly showing that {the} flow is anisotropic. {This is confirmed by
the simulation results. The error bars are relatively large.} {For ``weakly''
slipping surfaces,} the difference $b_{\mathrm{eff}}^{\parallel }-b_{\mathrm{%
eff}}^{\perp }$ is small compared to the slip lengths themselves, (of the
order of $\varepsilon ^{2}\ln \varepsilon$), and this is the reason for the
large {error} of the simulation {data}. {The simulation data agree with the
numerical results within the error. Nevertheless, the data suggest that they
lie systematically above the numerical results especially for larger
slipping phase fraction $\phi$. This is a consequence of the relatively
large Reynolds number. As discussed above, inertia effects primarily affect
the flow and effective slip length in the transverse configuration. Test
runs with larger shear rates were performed, and the deviations increased,
indicating that they presumably vanish in the Stokes limit.} Note that there has been recent (finite element method) work that
observed the decrease of superhydrophobic slip at large Reynolds numbers~\cite{cao.l:2012}, which is consistent with our results.

Now, we remark that further insight can be obtained from the above asymptotic results, Eqs.\ (\ref{b_par}) and (\ref{b_perp}), to predict
the dependence $b_{\mathrm{eff}}^{\parallel }-b_{\mathrm{eff}}^{\perp }$ on parameters of the texture:
\begin{equation}
\frac{b_{\mathrm{eff}}^{\parallel }-b_{\mathrm{eff}}^{\perp }}{L} \simeq -\frac{2b^{2}}{%
\pi L^2 }\left\{ \ln \left[ \frac{4\pi b}{L\sin \left( \pi \phi \right) }\right]
-\gamma \right\}   \label{b_dif}
\end{equation}
These values are also included in Figure \ref{fig:log_phi}. Eq.~(\ref{b_dif}), which can easily be handled,  demonstrates the
power of the asymptotic approach in deriving relevant expression for a difference in eigenvalues of the slip-length tensor.
In the limit of small $b/L$, the asymptotic expansion predicts correctly the positive
difference and enhanced anisotropy as the slip length increases. At larger $%
b/L$, deviations from the numerical results become larger due to the
increasing contribution from higher order terms. {At $b/L=0.034$, the
two-term prediction for the slip length difference is only in moderately
good agreement with the numerical data. For very low or very high coverage ($%
\phi \to 0$ or $\phi \to 1$), the agreement is not good at all, the theory
even predicts the wrong sign (not shown in Figure \ref{fig:log_phi}(b)). This is
consistent with our discussion in Section \ref{sec:e_s}, where we have
argued that the approximation must break down when the singular regions
associated with adjacent edges overlap.}
 {Also included in Figure \ref%
{fig:log_phi} is the result from the approximate expressions Eqs.\ (\ref%
{beff_par_largeH}) and (\ref{beff_ort_largeH}), which again shows
surprisingly good agreement with the numerical data} {over the whole range
of $\phi$.}

Our theory also allows one to quantify the drag force acting on a hydrophilic
sphere approaching a  ``weakly'' slipping stripes. It has been shown that such a geometry of configuration is equivalent to a sphere approaching  the imaginary smooth homogeneous isotropic surface shifted to a distance $s$ equal to the average of the eigenvalues of the effective slip-length tensor~\cite{asmolov_etal:2011}
\begin{equation}
s \simeq \frac{b_{\mathrm{eff}}^{\parallel}+b_{\mathrm{eff}}^{\perp}}{2} \label{shift1}
\end{equation}
The correction to a drag force due to superhydrophobic slip is then $f^{\ast} \simeq 1-s/h$~\cite{asmolov_etal:2011,lecoq.n:2004}.
By using Eqs.\ (\ref{b_par}) and (\ref{b_perp}) we could now easily relate $s$ to texture parameters by a simple analytical formula
\begin{equation}
\frac{s}{b} \simeq \phi + \frac{b}{\pi L} \left[ \ln \left( \frac{4 \pi^3 b^3}{L^3 \sin^3 (\pi \phi)}\right) - 3 \gamma \right]
 \label{shift2}
\end{equation}%
Eq.~(\ref{shift2}) can be also used in case of a plane decorated with shallow hydrophilic grooves, i.e. when the height of the texture, $e$, which should be used instead of $b$ then, is much smaller than $L$. This expression explains qualitatively recent experimental observations, where $s/e$ was found to be much smaller than $\phi$~\cite{mongruel.a:2013}. Unfortunately detailed quantitative comparison between the experimental results ~\cite{mongruel.a:2013} and our asymptotic predictions are impossible since the height of asperities in these experiments was not small
enough, $0.168 \le e/L \le 0.45$.

\begin{figure}[tbp]
\includegraphics[width=1.0\columnwidth]{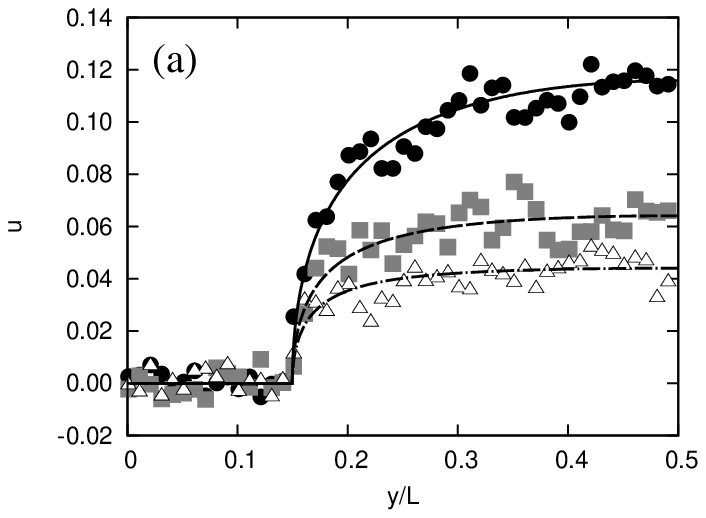} %
\includegraphics[width=1.0\columnwidth]{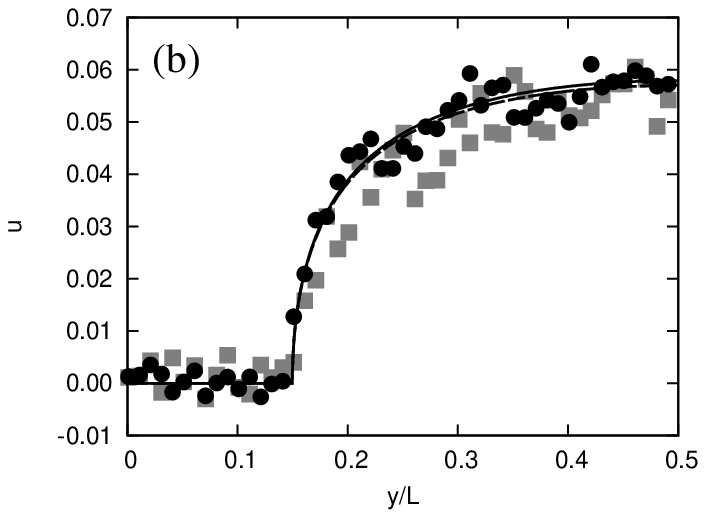}
\caption{(a) The longitudinal velocity along the wall for a texture with $%
\protect\phi=0.7$ and different slip lengths: $b/L=0.145$ (circles and solid
line), $b/L=0.071$ (squares and dashed line), and $b/L=0.047$ (triangles and
dash-dotted line). Symbols are simulation data and lines are numerical
results. (b) Comparison of the transverse velocity ($v[y,z=0,\protect\beta%
(y)]$) and longitudinal velocity for nearly double local slip length ($\frac{%
1}{2} u_d[y,z=0,2\protect\beta(y)]$, see Eq.~(\protect\ref{v0})). The
simulation and numerical results for longitudinal stripes of $b/L=0.147$ are
shown in circles and solid line, respectively. Squares and dashed line
correspond to transverse stripes of $b/L=0.071$. }
\label{fig:slip_vel}
\end{figure}

Finally, we consider the velocity at the wall near the edge of a heterogeneity. Figure \ref{fig:slip_vel}(a) presents the longitudinal velocity at the wall
for various $b/L$. {In the simulations, the slip velocity has been obtained
from an extrapolation procedure. Due to the small magnitude of the slip
velocity in comparison to the thermal fluctuation (order of 1 for $k_BT=1\epsilon$),
the data scatter very much. Much longer averaging times would be necessary
to improve the statistics.} The agreement of the exact numerical results and
DPD simulation data is again very good. The velocity distribution is not
smooth at the edge, {instead it rises according to a power law} on the
slipping area with {exponent} close to $\lambda=0.5$ {as} predicted in
Section \ref{sec:e_s}. In Figure \ref{fig:slip_vel}(b), we verify the
relation between the transverse and longitudinal velocities. {Eq.~(\ref{v0}%
) suggests that the local slip velocity above transverse stripes of slip
length $b$ should be identical to half of that above longitudinal stripes of
slip length $2b$. This is confirmed by the numerical results.} {The
simulation data, however, show deviations near the edge. This illustrates
the origin of the finite Reynolds number effects discussed above. In
simulations, the fluid is modelled as DPD particles with finite mass, and
abrupt changes of the transverse velocity are suppressed because of inertia.
Therefore the transverse velocity is smoothed out near the edge, showing a
smaller value compared to the numerical data.}

\section{Concluding Remarks}

\label{sec:conclusion}

{In conclusion,} we have investigated shear flow past ``weakly'' slipping
super-hydrophobic stripes, {focusing in particular on edge effects
associated with steplike discontinuities in the local slip length}. The
essential conclusion from our analysis is that {such step effects reduce the
effective slip below the surface-averaged value and induce anisotropy. In
practice, this means} that the flow does not align with the applied shear
stress. Thus, it {should} be possible to generate transverse hydrodynamic
phenomena (like in ~\cite{stone2004,vinogradova.oi:2011}) {even with such
``weakly'' slipping anisotropic textures}. This may also have relevance for
transverse electrokinetic phenomena~\cite%
{bahga:2009,vinogradova.oi:2011,Squires08,belyaev.av:2011a}. {As a side
remark,} our {analytical} result opens the possibility of solving
analytically many fundamental problems involving ``weakly'' slipping
heterogeneous surfaces, including hydrodynamic interactions.

Finally, we note that {even though} our discussion has been limited to
``weakly'' slipping heterogeneities, {our model is much more general. Every
result} in this work could be used for {describing} ``weakly'' rough or porous
surfaces since at large distances from the wall, the boundary condition at
the rough interface or fluid-porous interface may be approximated by a slip
model~\cite%
{Beavers-Joseph:1967,Taylor:1971,Richardson:1971,kunert-harting-07,kunert.c:2010,lecoq.n:2004}%
. In particular, our results allows one to interpret recent experiments with hydrophilic grooves, where even at small $e/L$ the model of ``average height'' significantly overestimated measured data~\cite{mongruel.a:2013}.

\appendix
\section{Divergence of the expansion (\ref{pow}), (\ref{bc_kbs}) for a discontinuous local slip length}\label{appA}

We consider periodic textures with $\beta \left( y\right) $ being an even
function, so that the slip length can be {expanded as a} cosine Fourier
series:%
\begin{equation}
\beta =\frac{\widetilde{\beta }^{0}}{2}+\sum\limits_{n=1}^{\infty }%
\widetilde{\beta }^{n}\cos \left( 2\pi ny\right) ,  \label{fourier}
\end{equation}%
\begin{equation}
\widetilde{\beta }^{n}=2\int_{-1/2}^{1/2}\beta \left( y\right) \cos \left(
2\pi ny\right) dy.  \label{bstar}
\end{equation}%
The expansions of the effective slip lengths {up to second order in $%
\varepsilon $ are then given by} \cite{kamrin_etal:2010}
\begin{eqnarray}
b_{\mathrm{eff}}^{||}/L &=&\varepsilon \frac{\widetilde{\beta }^{0}}{2}%
-\varepsilon ^{2}\pi \sum\limits_{n=1}^{\infty }n\left\vert \widetilde{\beta
}^{n}\right\vert ^{2},  \label{Kamr10a} \\
b_{\mathrm{eff}}^{\bot }/L &=&\varepsilon \frac{\widetilde{\beta }^{0}}{2}%
-\varepsilon ^{2}2\pi \sum\limits_{n=1}^{\infty }n\left\vert \widetilde{%
\beta }^{n}\right\vert ^{2}.  \label{Kamr10b}
\end{eqnarray}

The first-order terms are the isotropic part of the effective slip, Eq.\ (%
\ref{average}), since $\widetilde{\beta }^{0}=2\,\langle \beta (y)\rangle $.
The second-order terms, which can be neglected for a ``weakly'' slipping
patterns, are expected to introduce the influence of the surface structure,
and are responsible for the anisotropy of the flow.

{The }expansion (\ref{pow}), (\ref{bc_kbs}){\ implicitly assumes that the
infinite sums over $n$ in the higher-order expansion coefficients converge,
which implies that the Fourier series, Eq.\ (\ref{fourier}), can be
differentiated infinitely often with respect to $y$. In cases of
discontinuous slip, where $\beta (y)$ exhibits jumps, this is no longer
correct and the argument breaks down. }

The Fourier coefficients for the striped texture follow from Eq.\ (\ref%
{bstar})%
\begin{gather}
\widetilde{\beta }^{0}=2\phi ,  \label{Fb} \\
n>0:\quad \widetilde{\beta }^{n}=\frac{2\sin \left( \pi n\phi \right) }{\pi n%
}.  \notag
\end{gather}%
This implies that series in Eqs.\ (\ref{Kamr10a}) and (\ref{Kamr10b}),
\begin{equation}
\sum\limits_{n=1}^{\infty }n\left\vert \widetilde{\beta }^{n}\right\vert
^{2}=\frac{2}{\pi ^{2}}\sum\limits_{n=1}^{\infty }\frac{1-\cos \left( 2\pi
n\phi \right) }{n},  \label{sum1}
\end{equation}%
diverge, since their terms decay as $n^{-1}$ at $n\rightarrow \infty $
(large $n$ correspond to small length scales). The slow decay of $\left\vert
\widetilde{\beta }^{n}\right\vert ^{2}$ with $n$ and the divergence of the
series indicate that {the} expansion (\ref{pow}) does not resolve properly
the solution at small length scales.

\section{DPD Simulation}
\label{app:dpd}

To simulate the flow near striped superhydrophobic surfaces, we use a DPD version without conservative interactions \cite{Soddemann2003} and combine that with a tunable-slip method \cite{Smiatek2008} which allows one to implement arbitrary hydrodynamic boundary condition.

For two particles $i$ and $j$, we denote their relative displacement as $\mbox{\boldmath${r}$}_{ij}=\mbox{\boldmath${r}$}_i - \mbox{\boldmath${r}$}_j$, and their relative velocity $\mbox{\boldmath${v}$}_{ij}=\mbox{\boldmath${v}$}_i - \mbox{\boldmath${v}$}_j$.
We also introduce the distance between two particles $r_{ij}=  |\mbox{\boldmath${r}$}_{ij}|$ and the unit vector $\hat{\mbox{\boldmath${r}$}}_{ij}=  \mbox{\boldmath${r}$}_{ij}/r_{ij}$.
The basic DPD equations involve pair interaction between fluid particles.
The force exerted by particle $j$ {on} particle $i$ is given by
\begin{equation}
\mbox{\boldmath${F}$}_{ij}^{DPD} = \mbox{\boldmath${F}$}_{ij}^D + %
\mbox{\boldmath${F}$}_{ij}^R.
\end{equation}
The dissipative part $\mbox{\boldmath${F}$}_{ij}^D$ is proportional to the
relative velocity between two particles,
\begin{equation}  \label{eq:dpd_F_D}
\mbox{\boldmath${F}$}_{ij}^D = - \gamma_{DPD} \, \omega_D(r_{ij}) (%
\mbox{\boldmath${v}$}_{ij} \cdot \hat{\mbox{\boldmath${r}$}}_{ij}) \hat{%
\mbox{\boldmath${r}$}}_{ij}
\end{equation}
with a friction coefficient $\gamma_{DPD}$.
The weight function $\omega_D(r_{ij})$ is a monotonically decreasing function of $r_{ij}$, and vanishes at a given cutoff $r_c$.
The random force $\mbox{\boldmath${F}$}%
_{ij}^R$ {has} the form
\begin{equation}
\mbox{\boldmath${F}$}_{ij}^R = \sqrt{ 2 k_B T \gamma_{DPD} \,
\omega_D(r_{ij}) } \xi_{ij} \hat{\mbox{\boldmath${r}$}}_{ij},
\end{equation}
where $\xi_{ij}=\xi_{ji}$ are symmetric, but otherwise uncorrelated random
functions with zero mean and variance $\langle \xi_{ij}(t)
\xi_{ij}(t^{\prime}) \rangle = \delta(t-t^{\prime})$ (here $\delta(t)$ is
Dirac's delta function).
The magnitude of the stochastic contribution is related to the dissipative part by the fluctuation-dissipation theorem to ensure correct equilibrium statistics.
The pair forces between two particles satisfy Newton's third law, $\mbox{\boldmath${F}$}_{ij}=-\mbox{\boldmath${F}$}_{ji}$, and hence the momentum is conserved.
This leads to correct (\emph{i.e.} Navier-Stokes) long-time hydrodynamic behavior.

The wall interaction is introduced in the same spirit.
The force on particle $i$ from the channel wall is given by
\begin{equation}
\mbox{\boldmath${F}$}_i^{wall} = \mbox{\boldmath${F}$}_i^{WCA}+%
\mbox{\boldmath${F}$}_i^D + \mbox{\boldmath${F}$}_i^R .
\end{equation}
The first one is a repulsive interaction to prevent the fluid particles from penetrating the wall.
It can be written in terms of the gradient of a Weeks-Chandler-Andersen (WCA) potential \cite{WCA}
\begin{eqnarray}
\mbox{\boldmath${F}$}_i^{WCA} &=& - \nabla \cdot V(z),  \notag \\
V(z) &=& \left\{
\begin{array}{cl}
4 \epsilon [ (\frac{\sigma}{z})^{12} - (\frac{\sigma}{z})^6 + \frac{1}{4} ]
\quad & z < 2^{1/6} \sigma \\
0 & z \ge 2^{1/6} \sigma%
\end{array}
\right.
\end{eqnarray}
where $z$ is the distance between the fluid particle and the wall.
The energy and length units are denoted by $\varepsilon$ and $\sigma$, respectively.
The dissipative contribution is similar to Eq.\ (\ref{eq:dpd_F_D}), with the
velocity difference $\mbox{\boldmath${v}$}_{ij}$ replaced by the particle
velocity relative to the wall,
\begin{equation}
\mbox{\boldmath${F}$}_i^D = - \gamma_L \omega_L(z) (\mbox{\boldmath${v}$}_i
- \mbox{\boldmath${v}$}_{wall}).
\end{equation}
The parameter $\gamma_L$ characterizes the strength of the wall friction and
can be used to tune the value of the slip length.
For example, $\gamma_L=0$ corresponds to perfectly slippery wall, while a positive value of $\gamma_L$ leads to a finite slip length.
The weight function $\omega_L(z)$ is a monotonically decreasing function of $z$, and vanishes at a given cutoff $z_c$. The random term has the form
\begin{equation}
\mbox{\boldmath${F}$}_i^R = \sqrt{2 k_B T \gamma_L \omega_L(z)} \boldsymbol{%
\xi}_i ,
\end{equation}
where each component of $\boldsymbol{\xi}_i$ is an independent random variable function with zero mean and variance $\langle \xi_{i,\alpha}(t)\xi_{i,\alpha}(t^{\prime}) \rangle = \delta(t-t^{\prime})$.

The simulations are carried out using the open source simulation package
ESPResSo \cite{ESPResSo}.
We use a quadratically decaying weight function $\omega_D(r_{ij})$ and a linearly decaying weight function $\omega_L(z)$.
Table~\ref{tab:parameters} summarizes the simulation parameters.
The most important quantity is the slip length $b$, which can be estimated from the simulation parameters using an analytical formula due to \cite{Smiatek2008}.
However, the accuracy of the analytic {prediction is} not satisfactory for ``weakly'' slipping surfaces. We simulated plane Poiseuille and Couette flows with various $\gamma_L$ to obtain the slip length and the position of the hydrodynamic boundary (see \cite{Smiatek2008} for details).
Poiseuille flow was implemented by applying a constant body force of the order $10^{-4}
\epsilon/\sigma$ to all particles.

\begin{table*}[tbp]
\centering
\caption{Parameters used in the DPD simulations.}
\begin{tabular}{lp{0.4cm}l}
\\
\hline
Fluid density $\rho$ &  & $3.75 \sigma^{-3}$ \\
Friction coefficient for DPD interaction $\gamma_{DPD}$ &  & $5.0\sqrt{%
m\epsilon}/\sigma$ \\
Cutoff for DPD interaction $r_c$ &  & 1.0 $\sigma$ \\
Coefficient $\gamma_L$ for no-slip boundaries &  & $5.26 \sqrt{m\epsilon}%
/\sigma$ \\
Cutoff for wall interaction &  & 2.0 $\sigma$ \\
Coefficients $\gamma_L$ for slip boundaries &  & $\{0.5, \, 0.4, \, 0.3, \,
0.2, \, 0.1 \} \, \sqrt{m\epsilon}/\sigma$ \\
Corresponding slip lengths $b$ &  & $\{2.6, \, 3.4, \, 4.7, \, 7.1, \, 14.5
\} \, \pm 0.1 \, \sigma$ \\
Shear viscosity $\eta_s$ &  & $1.35 \pm 0.01 \sqrt{m\epsilon}/\sigma^2$ \\
Position of hydrodynamic boundary &  & $1.06 \pm 0.12 \sigma$ \\
Temperature $k_BT$ & & $1.0 \epsilon$ \\
\hline
\end{tabular}
\label{tab:parameters}
\end{table*}

In the simulations, the patterned surface has a stripe spacing of $L=100\sigma$.
The simulation box is a rectangular cuboid of size $20\sigma \times 100 \sigma \times 102\sigma$.
With a density of $3.0\,\sigma^{-3}$, this results in $6\times10^5$ particles in the simulation box.
Periodic boundary conditions are used in the $xy$ plane.
The integration time step is $0.01 \sqrt{m/\epsilon}\sigma$.
Due to the large system size, the time for reaching a steady state is also quite large (over $10^6$ time steps). For ``weakly'' slipping surfaces, the flow velocity near the wall is small compared to the thermal fluctuations; thus long simulation times are also required to obtain enough statistics.
In this work, velocity profiles have been averaged over $2\times10^5$ time steps.
The error bars of the simulation data are obtained from averaging over six independent runs.

%%%%%%%%%%%%%%%%%%%%%%%%%%%%%%%%%%%%%%%%%%%%%%%%%%%%%%%%%%%%%%

\begin{acknowledgments}
This research was supported by the RAS through its priority program `Assembly
and Investigation of Macromolecular Structures of New Generations', and by the
DFG through SFB-TR6 and SFB 985. The simulations were carried out using computational
resources at the John von Neumann Institute for Computing (NIC J\"ulich), the
High Performance Computing Center Stuttgart (HLRS) and Mainz University
(Mogon).
\end{acknowledgments}

\bibliographystyle{apsrev}
\bibliography{log}

\end{document}